\documentclass{IEEEtran}
\usepackage{graphicx}
\usepackage{bm}
\usepackage{cite}
\usepackage{pgfplots}
\usepackage{color}
\usepackage{algorithm}
\usepackage{amsmath,amssymb,amsthm,dsfont,mathrsfs}

\def\diag{{\mathrm{diag}}}

\def\A{{\mathbf{A}}}

\def\I{{\mathbf{I}}}

\def\y{{\mathbf{y}}}
\def\s{{\mathbf{s}}}

\def\R{{\mathbf{R}}}
\def\v{{\mathbf{v}}}
\def\z{{\mathbf{z}}}
\def\r{{\mathbf{r}}}

\def\g{{\mathbf{g}}}
\def\e{{\mathbf{e}}}

\begin{document}
%
\title{Disproving Sum-Difference Co-Array Property
}

\author{
\authorblockN{Jisheng~Dai\thanks{J. Dai is with the Department of Electronic  Engineering, Jiangsu University, Zhenjiang 212013, China (e-mail: jsdai@ujs.edu.cn).}}
and~\authorblockN{Hing Cheung So\thanks{H. C. So is with the Department of Electrical Engineering, City University of Hong Kong, Hong Kong, China (e-mail: hcso@ee.cityu.edu.hk).}}
}

\maketitle

\begin{abstract} The recently published paper by Gupta and Agrawal \cite{gupta2018design} exploited the sum-difference co-array (SDCA) to enhance the virtual aperture of sparse arrays. We argue that the key SDCA property established in  \cite{gupta2018design} requires a critical necessary and sufficient condition that is valid for a very rare case only.
\end{abstract}


\IEEEpeerreviewmaketitle

\section{Introduction}

Direction-of-arrival (DOA) estimation is one of the most important topics in array signal processing, and has attracted wide attention in radar, sonar and wireless communications. The classical subspace-type algorithms can resolve up
to $(N-1)$ sources for an $N$-element uniform linear array (ULA). To identify more sources than the number of sensors,
many sparse array geometries have been proposed in the last decade.

Recently, \cite{gupta2018design} claimed that the property of noncircular sources can be utilized to significantly  enhance the virtual aperture of sparse arrays.
Considering noncircular sources, its main idea is to combine the
difference co-array and sum co-array of the sparse array
configuration into a sum-difference co-array (SDCA), so that
the degree-of-freedom (DoF) can be substantially increased.
Then, \cite{gupta2018design} designed the sparse array  based on such SDCA, aiming to obtain the longest virtual ULA (VULA).
In this paper, we argue that the key SDCA property established in \cite{gupta2018design} needs a critical necessary and sufficient condition, and thus the corresponding array design criterion is generally invalid in practice.


\section{Review of SDCA in \cite{gupta2018design}}

Assume that $M$ narrow-band far-field sources impinge on an $N$-element linear array from distinct DOAs $\{\theta_1, \theta_2, \ldots, \theta_M\}$ with respect to the normal of the array.
Let the sensor positions be $\mathbb{L}=\{x_1, x_2,\ldots, x_N \}$. The array output $\y(k)=[y_1(k), y_2(k),\ldots, y_N(k)]^{\mathrm{T}}$ at the $k$th snapshot is given as:
\begin{align}\label{eq-model}
\y(k) = \A \s(k) + \v(k), ~~ k=1,2,\ldots, K,
\end{align}
where $\A=[\bm\alpha(\theta_1), \bm\alpha(\theta_2),\ldots, \bm\alpha(\theta_M) ]$,
$\bm\alpha(\theta_m)= [ e^{-j2\pi x_1 \sin(\theta_m)/\lambda} ,  e^{-j2\pi x_2 \sin(\theta_m)/\lambda} ,\ldots, e^{-j2\pi x_N \sin(\theta_m)/\lambda}     ]^{\mathrm{T}}  $, $\lambda$ denotes the wavelength,
$\s(k)=[s_1(k), s_2(k),\ldots, s_M(k)]^{\mathrm{T}}$ denotes the source vector, and $\v(k)=[v_1(k), v_2(k),\ldots, v_N(k)]^{\mathrm{T}}$ is the i.i.d. Gaussian noise vector.
For uncorrelated signals and noise, the covariance matrix of $\y(k)$ can be expressed as:
\begin{align}
\R_y= E\{\y(k) \y^{\mathrm{H}}(k) \}  =  \A \R_s \A^\mathrm{H}  +  \sigma_v^2 \I_N,
\end{align}
where $\R_s= \diag \{\g\}$ with $\diag\{\cdot\}$ being the diagonal operator, $\g=[g_1, g_2, \ldots, g_M]^{\mathrm{T}}$, $g_m= E\{ s_m(k) s_m^{*}(k) \} $, $(\cdot)^*$ denotes complex conjugate,
$\sigma_v^2$ is the noise variance, and $\I_N$ represents the $N\times N$ identity matrix.
On the other hand, taking the source noncircularity into account, the pseudo covariance matrix of $\y(k)$ is non-zero, which is calculated as:
\begin{align}
\bm\Gamma_y= E\{\y(k) \y^{\mathrm{T}}(k) \}  =  \A \bm\Gamma_s \A^\mathrm{T},
\end{align}
where $\bm\Gamma_s= \diag \{ \tilde{\g} \}$,  $\tilde{\g}=[\tilde{g}_1, \tilde{g}_2, \ldots, \tilde{g}_M]^\mathrm{T}$, and $\tilde{g}_m= E\{ s^2_m(k)  \} $.

To extend the array aperture, we may stack the  array output and its conjugate counterpart as $\z(k) =   [ \y(k)^\mathrm{T} , (\y^{*}(k))^\mathrm{T}   ]^\mathrm{T}$, and its covariance matrix is:
\begin{align}\label{eq-z}
\R_z=E\{\z(k) \z^{\mathrm{H}}(k) \}= \begin{bmatrix} \R_y &  \bm\Gamma_y\\
\bm\Gamma_y^{*} &   \R_y^*
\end{bmatrix}.
\end{align}
Rearranging $\R_z$ as $[ \R_y ,   \R_y^*,  \bm\Gamma_y,  \bm\Gamma_y^{*}  ]$,
followed by vectorization, we have:
\begin{align}
\r
\triangleq\begin{bmatrix}
\mathrm{vec}\{\R_y  \} \\
\mathrm{vec}\{\R_y^* \}\\
\mathrm{vec}\{\bm\Gamma_y \}\\
\mathrm{vec}\{\bm\Gamma_y^{*}  \}
\end{bmatrix}
=
\begin{bmatrix} (\A^*\odot\A) \g  + \sigma_v^2 \mathrm{vec}\{\I_N\}  \\
(\A\odot\A^*) \g  + \sigma_v^2 \mathrm{vec}\{\I_N\} \\
(\A\odot\A) \tilde{\g}\\
(\A^*\odot\A^* )\tilde{\g}^*
\end{bmatrix},
\end{align}
or, equivalently,
\begin{align}
\r= \begin{bmatrix}
\bm\Phi^{(1)} &&   \\
& \bm\Phi^{(2)} &\\
&&\bm\Phi^{(3)}
\end{bmatrix}
\begin{bmatrix}
\g\\
\tilde{\g}\\
\tilde{\g}^*
\end{bmatrix}
+ \e,  \label{eq-rg}
\end{align}
where $\mathrm{vec}\{\cdot\}$ is the  vectorization operator,
$\odot$ denotes the Khatri-Rao product,
$\bm\Phi^{(1)}= [(\A^*\odot\A)^{\mathrm{T}}, (\A\odot\A^*)^{\mathrm{T}} ]^{\mathrm{T}}$,
$\bm\Phi^{(2)}=\A\odot\A$, $\bm\Phi^{(3)}=\A^{*}\odot\A^{*}$,
$\e = [ (\sigma_v^2 \mathrm{vec}\{\I_{N}\})^\mathrm{T}, (\sigma_v^2 \mathrm{vec}\{\I_{N}\})^\mathrm{T}, \bm{0}^\mathrm{T}_{N^2\times 1}, \bm{0}_{N^2\times 1}^\mathrm{T}]^\mathrm{T}$, and $\bm{0}_{N^2\times 1}$
represents the zero column vector of length $N^2$.


The virtual steering vector corresponding to $\bm\Phi^{(1)}$ is determined by the difference co-array $\mathbb{D}_1$:
\begin{align} \label{eq-d}
\mathbb{D}^{(1)} =   \{  x_p - x_q   |~ x_p, x_q \in \mathbb{L}   \},
\end{align}
and the virtual steering vectors corresponding to $\bm\Phi^{(2)}$ and $\bm\Phi^{(3)}$ are chosen using the
positive-sum co-array $\mathbb{D}^{(2)}$ and negative-sum co-array $\mathbb{D}^{(3)}$, respectively:
\begin{align}
\mathbb{D}^{(2)} =&   \{   x_p + x_q   |~ x_p, x_q \in \mathbb{L}   \},     \label{eq-s+} \\
\mathbb{D}^{(3)} =&   \{  -(x_p + x_q)   |~ x_p, x_q \in \mathbb{L}   \}. \label{eq-s-}
\end{align}
In \cite{gupta2018design}, SDCA is defined as a union set of:
\begin{align}
\mathbb{S}= {\bigcup}_{i=1}^3 \mathbb{D}^{(i)}.  
\end{align}
Removing duplicate elements in $\mathbb{D}^{(i)}$s, we obtain  some truncated sets $\overline{\mathbb{D}}^{(i)}$s such that:
\begin{align}
&\overline{\mathbb{D}}^{(i)} \subseteq \mathbb{D}^{(i)}, ~\forall i, \\
&\overline{\mathbb{D}}^{(i)} \bigcap  \overline{\mathbb{D}}^{(j)} = \emptyset,~ \forall i \neq j,\\
&{\bigcup}_{i=1}^3 \overline{\mathbb{D}}^{(i)}  = \mathbb{S}.
\end{align}
The truncated data denoted as $\bar{\r}$ from (\ref{eq-rg}) can be written as:
\begin{align}
\bar{\r} =  \begin{bmatrix}
\bar{\bm\Phi}^{(1)} &&   \\
& \bar{\bm\Phi}^{(2)} &\\
&&\bar{\bm\Phi}^{(3)}
\end{bmatrix}
\begin{bmatrix}
\g\\
\tilde{\g}\\
\tilde{\g}^*
\end{bmatrix}
+ \bar{\e},
\end{align}
where $\bar{\bm\Phi}^{(i)}$ denotes a sub-matrix of $\bm\Phi^{(i)}$ with rows corresponding to the truncated set $\overline{\mathbb{D}}^{(i)} $ and $\bar{\e}$ denotes the truncated vector of $\e$.


The SDCA property in \cite{gupta2018design} claimed that $\bar{\r}$ can be viewed as the signal received by the SDCA, i.e.,
\begin{align}\label{eq-finalsd}
\begin{bmatrix}
\bar{\bm\Phi}^{(1)} &&   \\
& \bar{\bm\Phi}^{(2)} &\\
&&\bar{\bm\Phi}^{(3)}
\end{bmatrix}
\begin{bmatrix}
\g\\
\tilde{\g}\\
\tilde{\g}^*
\end{bmatrix}
\subset
\mathrm{span}\left\{\begin{bmatrix}
\bar{\bm\Phi}^{(1)}  \\
\bar{\bm\Phi}^{(2)}\\
\bar{\bm\Phi}^{(3)}
\end{bmatrix}\right\}
\end{align}
for any non-zero $\g$ or $\tilde\g$, where $\mathrm{span}\{\cdot\}$ denotes the subspace spanned by column vectors.

%
%

\section{Necessary and Sufficient Condition of SDCA}
The SDCA property plays a key role in designing the sparse array to achieve the largest VULA, but its proof was not provided in \cite{gupta2018design}. In the following, we argue that a critical necessary and sufficient condition is required.

{\bf Lemma 1:}  The SDCA property  holds if and only if
\begin{align}
g_m=\tilde{g}_m=\tilde{g}_m^*,~\forall m.
\end{align}

\begin{proof} 
\begin{itemize}
  \item $\Rightarrow$: The SDCA property  is applied to arbitrary number of sources (including $M=1$). If the $m$th source is considered in the data model only, (\ref{eq-finalsd}) will be reduced to:
   \begin{align}\notag
\begin{bmatrix}
\bar{\bm\Phi}_{:,m}^{(1)} &&   \\
& \bar{\bm\Phi}_{:,m}^{(2)} &\\
&&\bar{\bm\Phi}_{:,m}^{(3)}
\end{bmatrix}
\begin{bmatrix}
g_m\\
\tilde{g}_m\\
\tilde{g}_m^*
\end{bmatrix}
\subset
\mathrm{span}\left\{\begin{bmatrix}
\bar{\bm\Phi}_{:,m}^{(1)}  \\
\bar{\bm\Phi}_{:,m}^{(2)}\\
\bar{\bm\Phi}_{:,m}^{(3)}
\end{bmatrix}\right\}, \forall m,
\end{align}
where $(\cdot)_{:,m}$ denotes the $m$ column vector of a matrix.
For any non-zero $g_m$ or $\tilde{g}_m$, there always exists a non-zero scalar $\eta$ such that:
  \begin{align}\label{eq-L1-2}
  \begin{bmatrix}
    \bar{\bm\Phi}_{:,m}^{(1)} &&   \\
    & \bar{\bm\Phi}_{:,m}^{(2)} &\\
    &&\bar{\bm\Phi}_{:,m}^{(3)}
    \end{bmatrix}
    \begin{bmatrix}
    g_m\\
    \tilde{g}_m\\
    \tilde{g}_m^*
    \end{bmatrix}
    =
    \begin{bmatrix}
    \bar{\bm\Phi}_{:,m}^{(1)}  \\
    \bar{\bm\Phi}_{:,m}^{(2)}\\
    \bar{\bm\Phi}_{:,m}^{(3)}
    \end{bmatrix}\eta.
  \end{align}
  Equation  (\ref{eq-L1-2}) is equivalent to:
  \begin{align}
  \bar{\bm\Phi}_{:,m}^{(1)}  g_m=&  \bar{\bm\Phi}_{:,m}^{(1)} \eta,          \label{eq-L2-U1}\\
  \bar{\bm\Phi}_{:,m}^{(2)}  \tilde{g}_m=&  \bar{\bm\Phi}_{:,m}^{(2)} \eta,  \label{eq-L2-U2}\\
  \bar{\bm\Phi}_{:,m}^{(3)} \tilde{g}_m^*=&  \bar{\bm\Phi}_{:,m}^{(3)} \eta. \label{eq-L2-U3}
  \end{align}
  The equalities in (\ref{eq-L2-U1})--(\ref{eq-L2-U3}) indicate $g_m=\eta$,  $\tilde{g}_m= \eta$ and $\tilde{g}_m^*= \eta$, respectively. Therefore, $g_m=\tilde{g}_m=\tilde{g}^*_m$.

  \item $\Leftarrow$: If $g_m=\tilde{g}_m=\tilde{g}_m^*,~\forall m$, we obtain $\g=\tilde{\g}=\tilde{\g}^*$. Hence, we have:
  \begin{align}\label{eq-ff2}
\begin{bmatrix}
\bar{\bm\Phi}^{(1)} &&   \\
& \bar{\bm\Phi}^{(2)} &\\
&&\bar{\bm\Phi}^{(3)}
\end{bmatrix}
\begin{bmatrix}
\g\\
\tilde{\g}\\
\tilde{\g}^*
\end{bmatrix}
=\begin{bmatrix}
\bar{\bm\Phi}^{(1)}  \\
\bar{\bm\Phi}^{(2)}\\
\bar{\bm\Phi}^{(3)}
\end{bmatrix}\g,
\end{align}
which guarantees (\ref{eq-finalsd}) for any non-zero $\g$ or $\tilde\g$.

\end{itemize}
\end{proof}

With Lemma 1 and the definitions of $g_m$ and $\tilde g_m$, we obtain:
\begin{align}\label{eq-ress}
E\{ s_m(k) s_m^{*}(k) \} = E\{ s^2_m(k) \},~~\forall m.
\end{align}
Clearly, all $s_m(k)$s must be real numbers. At first glance, the SDCA property could hold true for any real-valued sources.
However, such real-valued constraint cannot be guaranteed even for real-valued sources.
It is worth noting that  the data model (\ref{eq-model}) is a simplified model, and the practical model is:
\begin{align}
\y(k) = \A  \begin{bmatrix} e^{j\phi_{1}} &&\\ &\ddots & \\ &&e^{j\phi_{M}}   \end{bmatrix}
\begin{bmatrix} s_1(k)\\ \vdots \\ s_M(k) \end{bmatrix}
 + \v(k), \label{model-prac}
\end{align}
where $\phi_{m}$ corresponds to the initial phase received by the first sensor for the $m$th source.
That is to say, the equivalent sources are $e^{ j\phi_{m}} s_m(k)$s instead of $s_m(k)$s.
The initial phases distort  the real-valued condition in (\ref{eq-ress}).
Since $\phi_{m}$s are randomly distributed, 
the SDCA property generally does not hold true
and is valid for a very rare case only. 


In the following simulation, we illustrate that ignoring the initial phases can significantly degrade the DOA estimation performance. Consider that a sparse array of 6 physical sensors located at $\mathbb{L}=\{ 0,\frac{\lambda}{2},\frac{2\lambda}{2},\frac{3\lambda}{2},\frac{10\lambda}{2},\frac{17\lambda}{2} \}$ \cite{gupta2018design}. Fig.~1 shows the root mean square error (RMSE) of DOA estimates (based on $1000$ Monte Carlo trials) versus signal-to-noise ratio (SNR), where two real-valued i.i.d. Gaussian sources uniformly come from intervals $[-20^\circ, -10^\circ]$ and $[20^\circ, 30^\circ]$ for each trial, and the number of snapshots is 200. The grid interval is chosen as $0.01^\circ$ for SS-MUSIC, and the smoothing window is maximally set to 40. It can be seen that SS-MUSIC works well for real-valued sources in the absence of the initial phases, but its performance substantially degrades if the practical model (\ref{model-prac}) is considered.

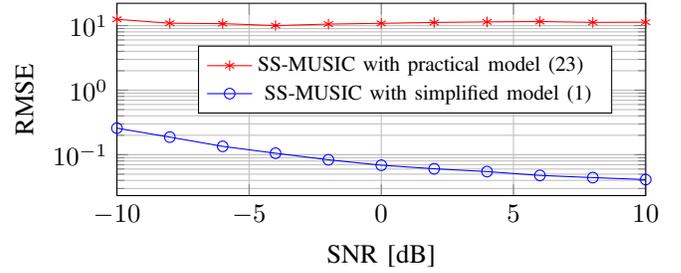
\begin{figure}
\center
\begin{tikzpicture}[scale=1]
\begin{semilogyaxis}[xlabel={SNR [dB]}, width=4.8cm,
ylabel={RMSE},grid=both,
legend style={at={(0.545,0.78),font=\footnotesize},
anchor=north,legend columns=1}, xmin=-10,xmax=10,x=10]
\addplot[mark=asterisk,red]  coordinates{
  (    -10,     12.5143)
  (     -8,     10.8856)
  (     -6,     10.7134)
  (     -4,     10.0220)
  (     -2,     10.5505)
  (      0,     10.8442)
  (      2,     11.1787)
  (      4,     11.4090)
  (      6,     11.5590)
  (      8,     11.2153)
  (     10,     11.2883)
};
\addplot[mark=o,blue]  coordinates{
  (    -10,     0.2579)
  (     -8,     0.1874)
  (     -6,     0.1347)
  (     -4,     0.1055)
  (     -2,     0.0835)
  (      0,     0.0688)
  (      2,     0.0606)
  (      4,     0.0548)
  (      6,     0.0478)
  (      8,     0.0442)
  (     10,     0.0412)
};
\legend{SS-MUSIC with practical model (\ref{model-prac})~~~, SS-MUSIC with simplified model (\ref{eq-model}) }
\end{semilogyaxis}
\end{tikzpicture}
\caption{RMSE of DOA estimates versus SNR for real-valued sources. 
 }
\end{figure}

\bibliographystyle{IEEEtran}
\bibliography{rootSBL}

\end{document}